\documentclass[a4paper, 10pt, twocolumn]{article}

\usepackage{amsfonts,amssymb,amsmath,amsthm,dsfont}
\usepackage{subfigure}
\usepackage{graphicx}
\usepackage[footnotesize]{caption}

\usepackage[top=1.5cm, left=1.5cm, right=1.5cm, bottom=1.5cm]{geometry}

\DeclareMathOperator*{\argmax}{argmax}

\newcommand{\ie}{\emph{i.e., }}
\newcommand{\eg}{\emph{e.g., }}

\newcommand{\Cr}{\mathds{C}}

\newcommand{\D}{\mathbf{D}}

\newcommand{\W}{\mathbf{W}}

\newcommand{\xv}{\mathbf{x}}
\newcommand{\hv}{\mathbf{h}}
\newcommand{\bv}{\mathbf{b}}
\newcommand{\av}{\mathbf{a}}

\newcommand{\yv}{\mathbf{y}}

\newcommand{\dv}{\mathbf{d}}

\newcommand{\wv}{\mathbf{w}}

\newcommand{\sv}{\mathbf{s}}
\newcommand{\rv}{\mathbf{r}}
\newcommand{\zv}{\mathbf{z}}

\newcommand{\Cs}{\mathcal{C}}

\newcommand{\Ns}{\mathcal{N}}

\title{Learning sparse structures for physics-inspired compressed sensing}

\author{Clément Dorffer, Thomas Paviet-Salomon,  Gilles Le Chenadec and Angélique Drémeau\\
\footnotesize ENSTA Bretagne and Lab-STICC, UMR CNRS 6285
}
\date{\empty} 

\renewenvironment{abstract}{\bf\small {\em\ Abstract---}}{}

\begin{document}

\maketitle

\begin{abstract} 
In underwater acoustics, shallow water environments act as modal dispersive waveguides when considering low-frequency sources. 
In this context, 
propagating signals can be described as a sum of few modal components, each of them propagating according to its own wavenumber.  Estimating these wavenumbers is of key interest to understand the propagating environment as well as the emitting source. To solve this problem, we proposed recently a Bayesian approach exploiting a sparsity-inforcing prior. When dealing with broadband sources, this model can be further improved by integrating the particular dependence linking the wavenumbers from one frequency to the other. In this contribution, we propose to resort to a new approach relying on a restricted Boltzmann machine, exploited as a generic structured sparsity-inforcing model. This model, derived from deep Bayesian networks, can indeed be efficiently learned on physically realistic simulated data using well-known and proven algorithms.
\end{abstract}

\section{Introduction}
\label{sec:introduction}

\begin{figure}[t!]
\begin{center}
	\includegraphics[width=0.6\columnwidth]{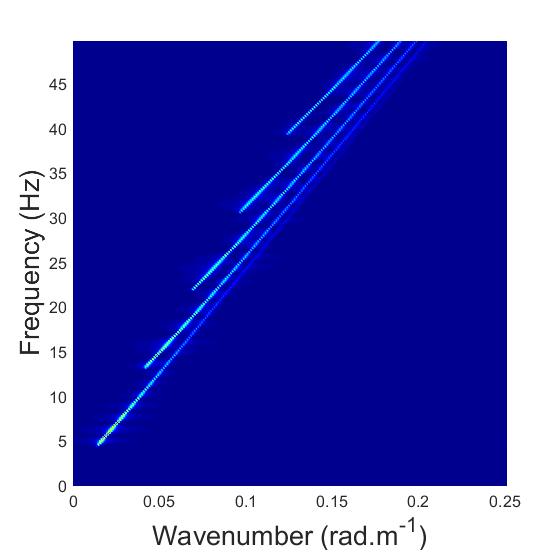}
	\caption{Illustration of a $f-k$ diagram obtained in a Pekeris waveguide.} \label{fig:illu}
\end{center}
\end{figure}

In underwater acoustics, ``shallow'' environments (about 100 m deep) behave like dispersive waveguides when considering low-frequency sources (below 250 Hz). An acoustic field received on an antenna is then classically described by a small set of modes propagating longitudinally according to their horizonal wavenumbers. The knowledge of these modes is of great importance for the characterization of the observation environment and, consequently, for the source localization. Among the different methods used to discriminate these modal components, the ``frequency-wavenumber'' ($f-k$) representations (see Fig. \ref{fig:illu}) allow a direct observation of the dispersion (\ie the frequency dependence) of the wavenumbers. Inherently conceivable for a horizontal array of sensors aligned with the source, they are particularly used in geophysics \cite{Amundsen1991}. Recent contributions have focused on the construction of $f-k$ diagrams by exploiting less constrained acquisition schemes, allowing their use in underwater acoustics.

Since propagation is described by a small number of modes, the use of sparsity-inforcing models seems appropriate. In fact, some contributions (see \eg \cite{Dremeau2017,Harley2015}) have proposed the use of the ``compressed sensing'' paradigm to estimate modal dispersion. However, if these methods prove to be relevant, we argue that they can be further improved by precisely integrating the dispersion relation linking the wavenumbers from one frequency to another into the estimation process of the $f-k$ diagram. To model this structured sparsity, we proposed in this paper to resort to a restricted Boltzmann machine (RBM), for which efficient learning algorithms exist.

The paper is organized as follows: the two next sections formalize the problem of wavenumber estimation and outline the prior models, including RBM, that are considered to take into account the physics of the problem. Section \ref{sec:third-section} describes the algorithm derived to solve the problem on the basis of these models, while Section \ref{sec:learning} discusses the learning of the RBM parameters. We conclude this abstract with a statement of intent regarding the experiments that will be conducted to validate the proposed approach and presented at the conference.

%

\section{Observation model}
\label{sec:first-section}
In shallow water, acoustic propagation is described by modal theory. According to the latter, when considering an emitting source $S(f)$, the signal received on an antenna at distance $r$ and at frequency $f$ can be described by
\begin{align}
y(f,r) \simeq QS(f) \sum_{m=1}^{M(f)}A_{m}(f) e^{irk_{rm}(f)}\label{eq:obs_an}
\end{align}
where $Q$ is a constant factor, $M(f)$ is the number of propagating modes at frequency $f$, $k_{rm}(f)$ (resp. $A_{m}(f)$) is the horizontal wavenumber (resp. modal amplitude) of the $m$-th mode. 

Adopting a discretized matrix formulation, Eq. \eqref{eq:obs_an} can be re-formulated as
\begin{align}
\yv= \D\zv + \wv\label{eq:obs_dis}
\end{align}
where $\yv\in\Cr^{LF}$ is the signal measured over the $L$-sensor antenna at all of the $F$ frequencies, $\D$ is a $(LF\times NF)$-dictionary of block-diagonal Fourier discrete atoms with $N$ the number of discretized points in the horizontal wavenumber domain, $\zv=[z_{1},\ldots,z_{NF}]^{T}$ is the vectorized $f-k$ diagram to estimate, and $\wv$ stands for an additive noise. 


\section{Prior assumptions}
\label{sec:second-section}
According to the modal theory, in shallow water environments and at low frequencies, 
the  vector $\zv$ has  few  non-zero  elements,  corresponding  to  the  propagating  modal wavenumbers.  
This sparsity constitutes important information on the $f-k$ diagram, that should be taken into account in the reconstruction procedure.  Several formulations of the corresponding sparse recovery problem can then be considered. In this paper, we focus on a Bayesian solution of the problem of wavenumber estimation. We thus consider the following prior models :  the noise $\wv$ is assumed to be circular Gaussian (denoted $\Cs\Ns$) with zero mean and variance $\sigma^{2}_{w}$, and $\zv$ is seen as the element-wise multiplication of two random vectors
\begin{align}
\zv=\sv \odot \xv
\end{align}
where $\xv$ is a multivariate Gaussian variable such that
\begin{align}
p(\xv) = \prod_{n=1}^{NF}p(x_{n}) \quad \text{with}\quad p(x_{n}) = \Cs\Ns(0,\sigma_{x}^{2}).
\end{align}
Classically, the variable $\sv$ stands for the support of the sparse representation and is assumed to obey a Bernoulli law. Here, we propose to change this prior into a so-called restricted Boltzmann machine (RBM), namely:
\begin{align}
p(\sv) = \sum_{\hv}p(\sv,\hv) \propto \sum_{\hv} \exp(\av^{T}\hv+\bv^{T}\sv+\sv^{T}\W\hv)
\end{align}
where $\hv$ is  a $P$-dimensional  binary  hidden  variable, $\av$ and $\bv$ stand for bias parameters and $\W$ can be seen as representative of the links between the coefficients in $\hv$ and $\sv$.  The  RBM  is the  building  block  of  ``deep  belief  networks'' \cite{Bengio2012}  and  has  recently sparked a surge of interest partly because of its huge representational power \cite{LeRoux2008} and the existence of efficient algorithms developed to train it (as the Contrastive Divergence (CD) \cite{Hinton2012}). All this makes them particularly well-suited for learning physically-based sparse structures.

\section{Deep structured SoBaP}
\label{sec:third-section}

Exploiting the model exposed in previous sections, we consider the following marginalized Maximum A Posteriori (MAP) estimation problem
\begin{align}
\hat{\sv} = \argmax_{\sv\in\lbrace 0,1\rbrace^{NF}} \log p(\sv|\yv),
\end{align}
where $p(\sv|\yv) = \int_{\xv}\sum_{\hv} p(\xv,\sv,\hv|\yv)d\xv$. To solve this problem, different sub-optimal techniques can be used. In the continuation of previous works \cite{Dremeau2012, Krzakala2012, Rangan2011}, we are interested here in the solutions brought by variational approaches, which aim to approximate the posterior distribution $p(\xv,\sv,\hv|\yv)$ by a distribution $q(\xv,\sv,\hv)$ leading to the minimum of the Kullback-Leibler divergence under specific sets of constraints. In particular, considering the factorization constraint
\begin{align}
q(\xv,\sv,\hv) &= \prod_{n=1}^{NF} q(x_{n},s_{n})\prod_{l=1}^{P}q(h_{l}) \\
&= \prod_{n=1}^{NF} q(x_{n}|s_{n})\; q(s_{n})\prod_{l=1}^{P}q(h_{l}),
\end{align}
we focus on a mean-field (MF) approximation, which can be in practice efficiently solved by an iterative algorithm, called ``variational Bayes EM algorithm'' \cite{Beal2003}. Particularized to our model, the method gives rise to the following iterative updates:
\begin{align}
	q(x_{n}|s_{n})&=\Ns(m(s_{n}),\Sigma(s_{n})),\\
	q(s_n)&\propto  \exp\left(s_n\bigg(b_n+\sum_l w_{nl}q(h_l=1)\bigg)\right)\nonumber\\
&\quad\sqrt{\Sigma(s_n)}\exp\left(\frac{1}{2}\frac{\vert m(s_n)\vert^2}{\Sigma(s_n)}\right)\label{eq-updt-qs}\\
	q(h_l)  &\propto \exp\negmedspace\left(\negmedspace h_l\bigg(a_l+\sum_n w_{nl}q(s_n=1)\negmedspace\bigg)\negmedspace\right)
\end{align}
where
\begin{align}
	\Sigma(s_{n})&=\frac{\sigma^2_{x}\sigma^2_w}{\sigma^2_w+ s_{n} \sigma^2_{x} \dv_n^T\dv_n},\\
	m(s_{n})&= s_{n} \frac{\sigma^2_{x}}{\sigma^2_w+ s_{n} \sigma^2_{x} \dv_n^T\dv_n} \langle\rv_{n}\rangle^T\dv_{n}, \nonumber\\
	\langle \rv_n \rangle&=\yv-\sum\nolimits_{j\neq n}q(s_j=1)\;m(s_j=1)\,\dv_j.\nonumber
\end{align}
The use of RBMs being a natural bridge towards deep networks, we refer to the above procedure as the ``Deep Structured Soft Bayesian Pursuit" (DSSoBaP). 
We propose here to apply it to the particular problem of $f-k$ diagram estimation.

\section{Learning RBM}
\label{sec:learning}

In \cite{LeRoux2008}, the authors  show that RBMs constitute universal approximators of discrete distributions. This characteristic makes them particularly interesting for generic but customizable structured sparse decomposition algorithms.
To particularise it to our problem, its parameters $\av$, $\bv$ and $\W$ have then to be trained to reflect at best the physical dispersion relation (see Fig. \ref{fig:illu}). This task can be efficiently performed by using the well-known ``Contrastive divergence'' algorithm \cite{Hinton2012}. 
This algorithm performs a (approximation of) maximum log-likelihood estimation of the RBM parameters with a very acceptable computation cost. It relies on a Markov Chain Monte Carlo approach in which the Markov chains are initialised at each step with samples from a training data set. 
The key point is then to produce a data set large enough to be representative of the different propagation environments. To this end, we will adopt a stepwise approach: RBM will first be trained on simulated data reproducing simple waveguides, such as Pekeris \cite{Jensen2011}, before considering more complex environmental models (multi-layer models). 

\section{On-going works}
\label{sec:conclusion}

To assess the performance of the proposed approach, two sets of experiments will be considered: $i)$ DSSoBaP will first be confronted to synthetic experiments, in accordance with the physics and of the same type as those used during the training step, $ii)$ it will then be applied to real data, acquired during a seismic campaign led by the Compagnie Générale de Géophysique \cite{Bonnel2010, Nicolas2003}. For both experiment sets, we will compare the results obtained by DSSoBaP with other state-of-the-art algorithms dealing with a simple sparse prior.

This work is on-going and will be presented at the conference, if accepted.



\end{document}